\documentclass[iop,revtex4]{emulateapj}
\usepackage{natbib, color}
\usepackage[normalem]{ulem}
\usepackage{epstopdf}
\usepackage{graphicx}
\definecolor{red}{rgb}{1.0,0.0,0.0}

\newcommand{\Mj}[1]{$M_\mathrm{Jup}$}
\newcommand{\Ms}[1]{$M_\odot$}

\begin{document}

%\tracingall

\accepted{}

\title{An Optical/near-infrared investigation of HD 100546 \lowercase{b} with the Gemini Planet Imager and MagAO}

\author{
Julien Rameau\altaffilmark{1},
Katherine B. Follette\altaffilmark{2,3},
Laurent Pueyo\altaffilmark{4},
Christian Marois \altaffilmark{5,6},
Bruce Macintosh\altaffilmark{2},
Maxwell Millar-Blanchaer\altaffilmark{7,8},
Jason J. Wang\altaffilmark{9},
David Vega\altaffilmark{10},
Ren\'{e} Doyon\altaffilmark{1},
David Lafreni\`{e}re\altaffilmark{1},
Eric L. Nielsen\altaffilmark{2,10},
%obs people
Vanessa Bailey\altaffilmark{2},
Jeffrey K. Chilcote\altaffilmark{11},
Laird M. Close\altaffilmark{12}
Thomas M. Esposito\altaffilmark{9},
Jared R. Males\altaffilmark{12,13},
Stanimir Metchev\altaffilmark{14,15},
Katie M. Morzinski\altaffilmark{12},
Jean-Baptiste Ruffio\altaffilmark{2},
Schuyler G. Wolff\altaffilmark{16,4},
%other
S. M. Ammons\altaffilmark{17},
Travis S. Barman\altaffilmark{18},
Joanna Bulger\altaffilmark{19},
Tara Cotten\altaffilmark{20},
Robert J. De Rosa\altaffilmark{9},
Gaspard Duchene\altaffilmark{9,21},
Michael P. Fitzgerald\altaffilmark{22},
Stephen Goodsell\altaffilmark{23,24},
James R. Graham\altaffilmark{9}, 
Alexandra Z. Greenbaum\altaffilmark{25},
Pascale Hibon\altaffilmark{26}, 
Li-Wei Hung\altaffilmark{22}, 
Patrick Ingraham\altaffilmark{27}, 
Paul Kalas\altaffilmark{9}, 
Quinn Konopacky\altaffilmark{28}, 
James E. Larkin\altaffilmark{7}, 
J{\'e}r{\^o}me Maire\altaffilmark{28}, 
Franck Marchis\altaffilmark{10}, 
Mark S. Marley\altaffilmark{29}, 
Rebecca Oppenheimer\altaffilmark{30}, 
David Palmer\altaffilmark{17}, 
Jennifer Patience\altaffilmark{31}, 
Marshall D. Perrin\altaffilmark{4},
Lisa Poyneer\altaffilmark{17}, 
Abhijith Rajan\altaffilmark{31}, 
Fredrik T. Rantakyr{\"o}\altaffilmark{24}, 
Dmitry Savransky\altaffilmark{32}, 
Adam C. Schneider\altaffilmark{31}, 
Anand Sivaramakrishnan\altaffilmark{4}, 
Inseok Song\altaffilmark{20}, 
Remi Soummer\altaffilmark{4},
Sandrine Thomas\altaffilmark{27}, 
J. Kent Wallace\altaffilmark{7}, 
Kimberly Ward-Duong\altaffilmark{31}, 
Sloane Wiktorowicz\altaffilmark{33} 
}

\altaffiltext{1}{Institut de Recherche sur les Exoplan\`{e}tes, D\'{e}partment de Physique, Universit\'{e} de Montr\'{e}al, Montr\'{e}al QC H3C 3J7, Canada}
\altaffiltext{2}{Kavli Institute for Particle Astrophysics and Cosmology, Stanford University, Stanford, CA 94305, USA}
\altaffiltext{3}{Physics and Astronomy Department, Amherst College, 21 Merrill Science Drive, Amherst, MA 01002, USA}
\altaffiltext{4}{Space Telescope Science Institute, 3700 San Martin Drive, Baltimore, MD 21218, USA}
\altaffiltext{5}{Department of Physics and Astronomy, University of Victoria, 3800 Finnerty Road, Victoria, BC, V8P 5C2, Canada}
\altaffiltext{6}{National Research Council of Canada Herzberg, 5071 West Saanich Road, Victoria, BC V9E 2E7, Canada}
\altaffiltext{7}{Jet Propulsion Laboratory, California Institute of Technology, 4800 Oak Grove Dr., Pasadena CA 91109, USA}
\altaffiltext{8}{NASA Hubble fellow}
\altaffiltext{9}{Astronomy Department, University of California, Berkeley, CA 94720, USA}
\altaffiltext{10}{SETI Institute, Carl Sagan Center, 189 Bernardo Avenue, Mountain View, CA 94043, USA}
\altaffiltext{11}{Dunlap Institute for Astronomy and Astrophysics, University of Toronto, Toronto, ON, M5S 3H4, Canada}
\altaffiltext{12}{Steward Observatory, 933 N. Cherry Avenue, University of Arizona, Tucson, AZ 85721, USA}
\altaffiltext{13}{NASA Sagan fellow}
\altaffiltext{14}{Department of Physics and Astronomy, Centre for Planetary Science and Exploration, The University of Western Ontario, London, ON N6A 3K7, Canada}
\altaffiltext{15}{Department of Physics and Astronomy, Stony Brook University, 100 Nicolls Road, Stony Brook, NY 11790, USA}
\altaffiltext{16}{Physics and Astronomy Department, Johns Hopkins University, Baltimore MD, 21218, USA}
\altaffiltext{17}{Lawrence Livermore National Laboratory, 7000 East Ave., Livermore, CA 94550, USA}
\altaffiltext{18}{Lunar and Planetary Lab, University of Arizona, Tucson, AZ, 85721, USA}
\altaffiltext{19}{Subaru Telescope, NAOJ, 650 North A’ohoku Place, Hilo, HI96720, USA}
\altaffiltext{20}{Department of Physics and Astronomy, University of Georgia,Athens, GA 30602, USA}
\altaffiltext{21}{Universit\'{e} Grenoble Alpes / CNRS, Institut de Plan\'{e}tologie et d'Astrophysique de Grenoble, 38000 Grenoble, France}
\altaffiltext{22}{Department of Physics and Astronomy, University of California Los Angeles, 430 Portola Plaza, Los Angeles, CA 90095, USA}
\altaffiltext{23}{Department of Physics, Durham University, Stockton Road, Durham, DH1 3LE, UK}
\altaffiltext{24}{Gemini Observatory, Casilla 603, La Serena, Chile}
\altaffiltext{25}{Department of Astronomy, University of Michigan, Ann Arbor, MI 48109, USA}
\altaffiltext{26}{European Southern Observatory, Alonso de Cordova 3107, Vitacura, Santiago, Chile}
\altaffiltext{27}{Large Synoptic Survey Telescope, 950N Cherry Av, Tucson, AZ 85719, USA}
\altaffiltext{28}{Center for Astrophysics and Space Science, University of California San Diego, La Jolla, CA 92093, USA}
\altaffiltext{29}{Space Science Division, NASA Ames Research Center, Mail Stop 245-3, Moffett Field CA 94035, USA}
\altaffiltext{30}{American Museum of Natural History, Department of Astrophysics, New York, NY 10024, USA}
\altaffiltext{31}{School of Earth and Space Exploration, Arizona State University, PO Box 871404, Tempe, AZ 85287, USA}
\altaffiltext{32}{Sibley School of Mechanical and Aerospace Engineering, Cornell University, Ithaca, NY 14853, USA}
\altaffiltext{33}{The Aerospace Corporation, 2310 E. El Segundo Blvd., El Segundo, CA 90245}

\begin{abstract}
We present {\it H} band spectroscopic and {\it H}$\alpha$ photometric observations of HD 100546 obtained with the Gemini Planet Imager and the Magellan Visible AO camera. We detect {\it H} band emission at the location of the protoplanet HD\,100546\,b, but show that choice of data processing parameters strongly affects the morphology of this source. It appears point-like in some aggressive reductions, but rejoins an extended disk structure in the majority of the others. Furthermore, we demonstrate that this emission appears stationary on a timescale of 4.6 yrs, inconsistent at the $2\sigma$ level with a Keplerian clockwise orbit at $59$ au in the disk plane. The {\it H} band spectrum of the emission is inconsistent with any type of low effective temperature object or accreting protoplanetary disk. It strongly suggests a scattered light origin, as it is consistent with the spectrum of the star and the spectra extracted at other locations in the disk. A non-detection at the $5\sigma$ level of HD\,100546\,b in differential {\it H}$\alpha$ imaging places an upper limit, assuming the protoplanet lies in a gap free of extinction, on the accretion luminosity of $1.7\times10^{-4}~\mathrm{L}_\odot$ and $\mathrm{M}\dot\mathrm{M}< 6.3\times10^{-7} ~\mathrm{M}_\mathrm{Jup}^2.\mathrm{yr}^{-1}$ for 1 $\mathrm{R}_\mathrm{Jup}$. These limits are comparable to the accretion luminosity and accretion rate of T-Tauri stars or LkCa 15 b. Taken together, these lines of evidence suggest that the {\it H} band source at the location of HD\,100546\,b is not emitted by a planetary photosphere or an accreting circumplanetary disk but is a disk feature enhanced by the PSF subtraction process. This non-detection is consistent with the non-detection in the {\it K} band reported in an earlier study but does not exclude the possibility that HD\,100546\,b is deeply embedded. 
    
\end{abstract}


\section{Introduction}
The evolution of circumstellar disks is driven by accretion/ejection \citep{lynden74,armitage11,alexander13}, photoevaporation \citep{Johnstone93,armitage11}, dust growth, and planetary formation \citep{crida07,dullemond09, testi14, espaillat14}. These mechanisms act simultaneously, albeit with different efficiencies, over timescales of up to a few million years. They are responsible for dramatic changes in disk properties and observational signatures, and might ultimately create the centrally-cleared disk cavities that are the signature feature of the transitional disk subclass.

The transitional disk phase was first identified by near infrared flux deficits in the spectral energy distributions of young objects compared to those of class II objects \citep{strom89}, and was later interpreted as the formation of an optically thin gap or cavity at the disk center \citep{espaillat11}. Resolved images of transition disks in a broad range of wavelengths have since revealed rich spatial structures in both dust and gas tracers, including: gaps, cavities, spirals, and vortices \citep[e.g.,][]{fukagawa06,andrews11,muto12,debes13,casassus13,vandermarel13,quanz13,follette15}. Several recent observational results have refuted photoevaporation and viscous evolution as the main mechanisms responsible for transition disks \citep{pinilla12,garufi13,vandermarel13}, as they cannot reproduce asymmetric structures \citep{owen12}. Thus, planetary formation is emerging as a very likely mechanism for the creation of observed transition disk structures \citep{crida07,dodson:2012,birnstiel13,espaillat14,Zhu:2015}, in particular the dust segregation as seen in different wavelength regimes \citep[e.g.,][]{Pinilla:2015b}.

Observationally, protoplanetary disks are important in that they allow us to directly observe the first stages of planet formation. Recent discoveries of a small number of low-mass companions inside transition disk gaps \citep{kraus12,biller12,close14,Sallum:2015ej} are aiding in our understanding of planet formation. 

The detection of point-like emissions near observed disk features such as inner rims \citep[e.g.,][]{huelamo11} require cautious interpretation, as they could be image artifacts stemming from the circumstellar disk and not actual companions \citep{olofsson13,Thalmann:2016}. The cospatial nature of planets and disks therefore makes the distinction between disk structures and real companions nontrivial. This is particularly true when searching for self-luminous companions embedded in an optically thick circumstellar disk, and is especially complicated when using adaptive optics (AO) imaging with the Angular Differential Imaging technique \citep[ADI,][]{Marois:2006df}. During an ADI sequence, the field of view rotates with respect to the detector while the optics of the telescope and the instrument remain stable. Any rotating astrophysical source can then be separated from the fixed point spread function (PSF) and from quasi-static speckles. This observing mode, and the associated data reduction process, induces flux loss of the astrophysical signal and produces morphological hence measurement biases. For extended and inclined structures like disks, PSF subtraction techniques may also create asymmetries and blobs \citep{Milli:2012}, which can mimic point sources and be interpreted as embedded companions. 

HD 100546 (spectral type of B9V, mass of $2.4\pm0.1$ \Ms~, distance of $109\pm 4$ pc, age of $5-10$ Myr, \citealt{levenhagen06,vanderancker97,gaia16,Guimaraes06,vanderancker97}) hosts an extensively studied transition disk that extends out to $300$ au and exhibits a number of peculiar and potentially planet-induced morphological features, including: a gap, spiral arms, and asymmetries. \citep[e.g.,][]{Augereau:2001,liu03,grady05,ardilla07,benisty10,quanz11,mulders13,boccaletti13,avenhaus14,panic14,pineda14,pinilla15,Garufi:2016}. A protoplanet, HD 100546 b, has been detected by AO direct imaging at L\,' ($3.8$ $\micron$) and M\,' ($4.8$ $\micron$) with VLT/NaCo, and is embedded in the Northen part of the disk at a physical separation\footnote{The original separation in the discovery paper was $53$ au but it has been updated and propagated throughout the paper based on the new \textit{Gaia} parallax measurement \citep{gaia16}.} of $59\pm2$ au \citep{Quanz:2013,Quanz:2015}. Its mass is not well constrained ($5-15$ \Mj~~) because the origin of the emission is unclear (planet photosphere or circumplanetary disk) and also because the age of the star and of the (perhaps much younger) protoplanet is not well determined. However, the emission has a black body effective temperature of $\approx900$ K, and a luminosity of $\approx2.6\times10^{-4}$ L$_\odot$ \citep{Quanz:2015}. Analysis of its morphology showed that it is composed of a point source (the protoplanet and an unresolved circumplanetary disk) surrounded by a warm component of the surrounding circumstellar disk \citep{Quanz:2015}. However, beyond the extent of the resolved emission, the rest of the field of view is clear of any disk emission, and HD 100546 b lies in an weakly polarized region of the disk \citep{Quanz:2013, avenhaus14}, favoring a protoplanet scenario compared to a disk hot spot. Recently, with shorter wavelength ($\lambda_c=1.64$ $\micron$), a detection of HD 100546 b was reported with the Gemini Planet Imager \citep[GPI,][]{Macintosh:2014js} along a very bright disk arc extending from the protoplanet location to the coronagraph edge \citep{Currie:2015}.

To follow-up this peculiar system, we obtain considerably longer dataset with large field rotation as previously reported. HD100546 was observed with the GPI Instrument at Gemini South as part of the Gemini Planet Imager Survey (GPIES) and with the Visible AO (VisAO) camera at the Magellan Clay Telescope at Las Campanas Observatory. In this paper, we present images of a source detected at the location of HD 100546 b in {\it H} band but not in {\it H}$\alpha$. We describe four independent arguments assessing the nature of this source based on its morphology (Sec. \ref{sec:planet}), astrometry (Sec. \ref{sec:spectro}), {\it H} band spectrum (Sec. \ref{sec:spectrum}), and accretion luminosity (Sec. \ref{sec:accretion}). Numerous disk structures are also resolved in these data but their description and interpretation are presented in a companion paper (Folette et al., submitted).

\section{Observing strategy and image processing}
\label{sec:obs}

%\begin{deluxetable*}{lccccccccc}[th]
%\tablecaption{Observing log of HD 100546}
%\tablewidth{0pt}
%\tablehead{
%\colhead{Instrument} & \colhead{UT-date} & \colhead{Camera} & \colhead{Mode} & \colhead{Filter} & \colhead{Exposure time} & \colhead{N$_\mathrm{exp}$} & \colhead{FoV rotation} & \colhead{$\langle$Airmass$\rangle$} &   \colhead{$\langle\varpi\rangle$} \\
%\colhead{} & \colhead{} & \colhead{}    & \colhead{}  & \colhead{} & \colhead{(s)}  & \colhead{} &\colhead{(deg)} & \colhead{} &  \colhead{($\,''$)}}
%\startdata
%GPI & 2016/02/27 & IFS & Spec. & {\it H}& 60 & 120 & 51.6 & 1.36 & 0.87  \\
%\hline
%MagAO & 2014/04/12 & VisAO & SDI & {\it H}$\alpha$/Cont. & 2.27 & 4939 & 72 & -- & 0.58
%\enddata
%\label{tab:log}
%\tablecomments{The airmass and the DIMM seeing $\varpi$ averaged over the observing sequence when available. "Cont." stands for a filter in the nearby continuum of the {\it H}$\alpha$ line.}
%\end{deluxetable*}

\begin{table*}[th]
\begin{center}
\caption{Observing log of HD 100546.}
\label{tab:log}
\begin{tabular}{lccccccccc}
\noalign{\smallskip}
\noalign{\smallskip}\hline
\noalign{\smallskip}\hline  \noalign{\smallskip}
Instrument & UT-date & Camera & Mode & Filter & Exposure tine & N$_\mathrm{exp}$ & FoV rotation & $\langle$Airmass$\rangle$ & $\langle\varpi\rangle$ \\
 & & & & & (s) & & (deg) & & $\arcsec$ \\
 \noalign{\smallskip}\hline                  \noalign{\smallskip}
 GPI & 2016/02/27 & IFS & Spec. & {\it H}& 60 & 120 & 51.6 & 1.36 & 0.87  \\
MagAO & 2014/04/12 & VisAO & SDI & {\it H}$\alpha$/Cont. & 2.27 & 4939 & 72 & -- & 0.58 \\
  \noalign{\smallskip}\hline  \noalign{\smallskip}
%\tablenotetext{\small The airmass and the DIMM seeing $\varpi$ averaged over the observing sequence when available. "Cont." stands for a filter in the nearby continuum of the {\it H}$\alpha$ line.}
\end{tabular}
\begin{tabular}{p{0.85\textwidth}}
\small The airmass and the DIMM seeing $\varpi$ averaged over the observing sequence when available. "Cont." stands for a filter in the nearby continuum of the {\it H}$\alpha$ line.
\end{tabular}
\end{center}
\end{table*}

\subsection{Gemini Planet Imager}
\begin{figure*}[th]
    \centering
    \includegraphics[width=\textwidth]{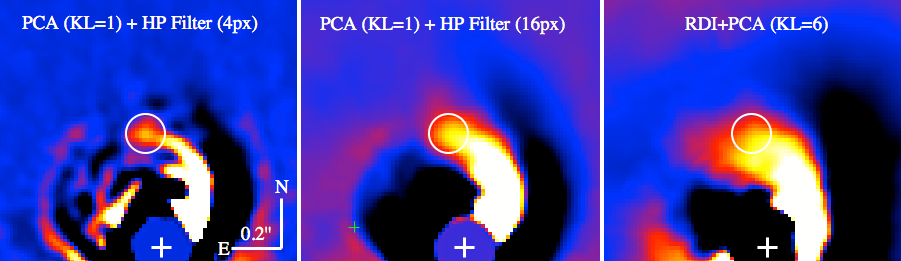}
    \caption{GPI $H$ band images of HD 100546 after PSF subtraction in ADI mode (left, middle) and RDI mode (right) with the PCA algorithm. The expected location of HD 100546 b is highlighted by the white circle in all three panels but is point-like only when using aggressive high-pass filter (left). The images are centered on the source at the location of HD 100546 b, and the position of the star is indicated by a plus sign. In each image, the central mask is numerical and all intensity scales are linear.}
    \label{fig:gpiimage}
\end{figure*}

HD 100546 was observed on UT 2016-02-27 (program ID: GS-2015A-Q-501). These data are fundamentally new observations that gave us the capacity to present the first spectrum in {\it H} band. A total of $120\times60$ second IFU images were taken in $H$ band ($1.49-1.79$ $\micron, \lambda/\Delta\lambda\approx45$) in coronographic mode using ADI. A total of $51.6$ deg of field rotation (FoV) was obtained over the sequence. Observing conditions were stable and good with an average DIMM seeing of $0.8\arcsec$ and coherence time of $4$ ms. The observing parameters are reported in Table \ref{tab:log}. Observations of an argon arc-lamp were acquired at the target elevation immediately prior to the sequence and were combined with reference arcs taken at zenith to calculate flexure compensation \citep{Wolff:2014}. The $\theta_1$ Ori field was observed to calibrate the platescale ($14.166\pm0.007$ mas.pixel$^{-1}$) and the correction to the detector position angle of ADI observations between the North and the vertical axis ($0.1\pm0.13$ deg, \citealt{Konopacky:2014}).

Another set of observations of HD 100546 was obtained on UT 2014-12-17 (GS-2014B-Q-500) but with only $12.9$ deg of FoV rotation. Therefore, this dataset was not used in the present paper.

The raw 2-D images were initially processed with the GPI Data Reduction Pipeline v1.3.0 \citep[DRP,][]{Perrin:2014} which performs dark subtraction, bad pixel identification and removal, instrument flexure correction, extraction of spectra to create 3-D $(x,y,\lambda)$ data cubes \citep{Maire:2014gs}, interpolation over a common wavelength axis, and distortion correction as measured with a pinhole mask \citep{Konopacky:2014}. The positions of the four satellite spots - attenuated replicas of the central occulted star - at each wavelength in each data cube were also measured with the DRP \citep{Wang:2014}. The barycenter of each set of spots was then used to align all the images to the location of the star.

The reduced data cubes were then processed to subtract the stellar point spread function (PSF). Large-scale structures (e.g., temporal variation of the residual turbulence affecting the stellar halo and background) slowly varying from one frame to another are not fully subtracted by ADI algorithms and must be removed before the PSF subtraction. The data were therefore high-pass filtered. An apodized Fourier filter was used to attenuate, following a Hanning profile, spatial frequencies lower than a parametrized cutoff. Different sizes were tested from 4 pixels up to 16 pixels. To subtract the PSF, three ADI algorithms were applied and the results were compared: the standard cADI \citep{Marois:2006df}, LOCI (\citealt{Lafreniere:2007};$dr=5$ pixels, $N_A=500$ full width at half maximum, FWHM, $3.6$ pixels in $H$ band, $g=1$, and $N_\delta=0.75$ FWHM), and PCA (\citealt{Soummer:2012, Amara:2012}; 1-3 Karhunen-Lo\`eve (KL) modes kept, a single region from $3$ to $100$ pixels in radius). For each algorithm, the residual frames were rotated to align North with the vertical axis and combined with a trimmed mean (10\%) in both temporal and spectral directions, resulting in a single image.

 The data were also processed using the Reference Differential Imaging (RDI) technique, which is much less susceptible to self-subtraction of disk features, and was not available in the previous works since we benefit from the GPIES reference library. First, the data were processed with \textsc{pyKLIP} \citep{Wang:2015}, a \textsc{python} implementation of the KLIP algorithm, to produce a RDI PSF subtracted broad-band image. Second, since the RDI \textsc{pyKLIP} pipeline does not handle spectral datacubes, the data were processed with TLOCI to produce 37 RDI PSF subtracted spectral images. For the first reduction, a library of reference images was created from all the 6265 individual spectroscopic $H$ band observations obtained with GPI as part of GPIES,at the time were the data were reduced, but rejected targets with known disks and/or companions to ensure star subtraction only (\citealt{Draper:2016}, Millar-Blanchaer et al. in prep). Each observation was reduced with the GPI DRP as described previously and collapsed over the wavelengths to populate a homogeneous library of broadband images. HD 100546 data cubes were processed through \textsc{pyKLIP},  using the 1000 most correlated reference images in a 10-70 pixel annulus to avoid the mask edges and the satellite spots. This process produces a broad-band image subtracted from the star PSF that can be compared to the ADI reduced images. For the second reduction, the TLOCI quicklook pipeline was used. The library and the HD 100546 datacubes were spatially scaled to align speckles based on the satellite spot locations in each wavelength channel and flux normalized with the spots. Each reference library datacube was median-combined to create 426 achromatic reference images. For each frame of HD 100546, the 20 most correlated images of the library were median-combined and subtracted to remove the star PSF. No high-pass filter was applied to preserve large scale disk structures. The RDI-processed images of HD 100546 were scaled back to the original spatial resolution, and rotated to align the North with the vertical axis. We ended up with a 37 channel RDI-reduced datacube that was used to extract the spectrum of the disk.

\subsection{MagAO}
Magellan Adaptive Optics observations of HD100546 were taken on UT 2014-04-12 in better than median conditions (average seeing 0$\farcs$58). Total integration time for the image set was 187 minutes, and individual exposures were limited to 2.3 seconds to minimize saturation effects. The data were taken in Angular Differential Imaging mode, and a total of 72$^{\circ}$ of rotation were achieved (see Table \ref{tab:log}).

We utilized the Simultaneous Differential Imaging (SDI) mode of MagAO's visible light camera VisAO to image the disk simultaneously in {\it H}$\alpha$ and in the neighboring continuum as part of the Giant Accreting Protoplanet Survey (GAPlanetS, Follette et al., in prep). Under this mode, the continuum channel serves as a sensitive and simultaneous probe of both the stellar PSF and any scattered light features.

The SDI data reduction and analysis procedures are described in detail in the companion paper to this study (Follette et al., accepted). Briefly, a custom \textsc{IDL} pipeline performed bias subtraction, flat field correction, splitting in two channels corresponding to {\it H}$\alpha$ and continuum and star registration. The PSF in each channel was estimated and removed using \textsc{pyKLIP} with 30 pixel wide annuli and a movement criterion of 12 pixels, a paramater silimar to $N_\delta$ in LOCI and defined by the number of pixels a source would move azimuthally and radially due to ADI and SDI. The non-coronagraphic images were saturated out at 8 pixels in radius, and so a 10 pixel radius mask was applied to all images. The region near the AO control radius, which is dominated by quasi-static speckles, was also masked. SDI images were created by scaling and subtracting each continuum frame from its {\it H}$\alpha$ counterpart, and these were then processed with \textsc{pyKLIP}. SDI processing serves to remove both PSF artifacts common to both channels and any scattered light structures present in the image; excess emission in the {\it H}$\alpha$ channel is thus preserved.

Contrast curves were also generated. The SDI residual image was convolved with an estimate of the PSF, the noise was measured in annuli of 1 FWHM in width and corrected for small sample statistics \citep{Mawet:2014}. Algorithm throughput at the location of any accreting source was computed by injecting fake planets into the raw {\it H}$\alpha$ datasets and processing through SDI and \textsc{pyKLIP}. We measure only the contrast limit for an {\it H}$\alpha$ excess source like LkCa15\,b \citep{Sallum:2015ej}. Therefore, we are only looking for accreting objects considering that any forming planet surrounded by a massive circumstellar disk would be heavily fed by gas. We refer to the paper of Follette et al. (accepted) for additional details and to Section \ref{sec:accretion} for a discussion of several hypotheses.

\section{Morphology at the position of HD 100546 \lowercase{b} }
\label{sec:planet}

 PCA images, with two different high pass filter sizes, and RDI images are shown in Figure \ref{fig:gpiimage}. The key feature is a smooth disk arc that is resolved to the North West of the star, which is also seen in the SPHERE IRDIS image as the ``West wing" \citep{Garufi:2016}. Moreover, a point-source lying along this structure is revealed at the expected location of HD\,100546\,b with aggressive reduction parameters. We present several arguments regarding the nature of this source in the most aggressive reductions.

No negative lobes flank the source although they are inherent to ADI processing of point sources. They are created by the point source being partly present at different parallactic angles in the reference PSF images built with ADI algorithms \citep{Marois:2006df,Lafreniere:2007,Soummer:2012}.

When using a larger high-pass filter as displayed in the middle panel of Figure \ref{fig:gpiimage}, the source rejoins the arc and appears as its natural extension. This behavior is even more evident with RDI as shown in the right panel, the disk being extended to the East with the brightest part being consistent with the arc seen in ADI.

Sharp disk features are highlighted by both the ADI process, which acts as a frequency filter, and by further high pass filtering. The more aggressive the filter and ADI algorithm parameters are, the more prominent these sharp structures appear. \citet{Milli:2012} demonstrated similar effects on observations of disks with inclination less than $50$ deg, like HD 100546: clumps appear and can lead to misinterpretation of the data. In our data, this is most easily seen in the spiral arm resolved on the East side of the disk that extends nearly up to the location of HD 100546\,b, clearly seen in the left panel of Figure \ref{fig:gpiimage}. The very end of this spiral is broken into blobs by aggressive data processing, and just like our source, these blobs are not robust across algorithms and algorithmic parameters.

\begin{figure}[th]
    \centering
    \includegraphics[width=0.5\textwidth]{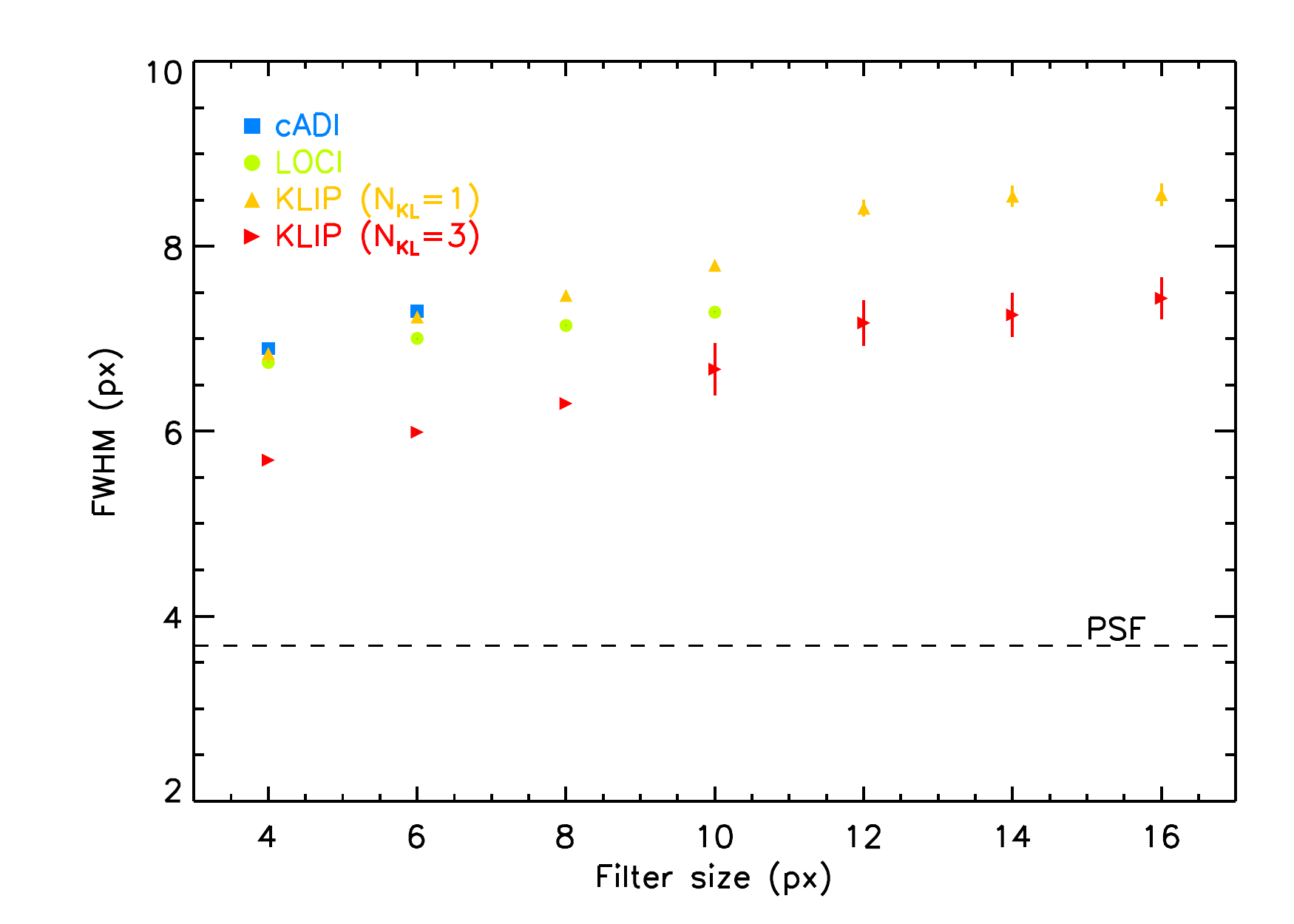}
    \caption{Influence of high-pass filter size on the width of the source at the expected location of HD 100546 b in the GPI $H$ band image with different ADI algorithms. The cADI and LOCI measurements end with filter sizes of $6$ and $10$ pixels respectively since the source was no longer independent from the rest of the arc. Error bars are plotted but are often smaller than the size of the symbols.}
    \label{fig:fwhm}
\end{figure}

Quantitatively, we measured the FWHM of the source as a function of the filter size from $16$ to $4$ pixels, i.e., from $4.5$ to higher than $1$ GPI $H$ band FWHM, for all three ADI algorithms. A Gaussian function was fit to the position, size, and flux of the source in each residual image in a wedge of $3\times3$ GPI $H$ band FWHM centered on the position of the source. Errors were estimated by varying the size of the wedge. To make sure that the fitting process does not subtract both source and disk, the procedure adopted by \citet{Quanz:2015} was followed. Briefly, the fitting is done such that the subtraction of the best fit function to the data does not over-subtract the underlying disk arc. Briefly, the determinant of the Hessian matrix computed at that location would be negative due to an oversubtraction, showing that the surrounding disk has been affected by the process. The result of the analysis is shown in Figure \ref{fig:fwhm}. The size of the source strongly depends on the filter size, becoming more compact with a smaller filter. Even for the smallest filter, the source is still larger than a point source FWHM. 

While it is possible that a true object embedded in the extended disk might be revealed with this variety of aggressive processing, the morphology of the {\it H} band source at the location of HD100546 b, together with the impact of filter size and reduction algorithm, are most consistent with a disk artifact rather than a standalone point source. Note however that this argument does not formally rule out the presence of an astrophysical object, which is why we then pursue other lines of inquiry.

\section{Physical properties at the position of HD 100546 \lowercase{b} }
\label{sec:spectro}

If the source detected within our GPI {\it H} band data is consistent with HD\,100546\,b seen in L\,' and M\,', it should have the characteristics of a protoplanet (or protoplanet plus circumplanetary disk) with effective temperature of $\approx900$ {\it K} orbiting close to 60 au from its host star as derived by \citet{Quanz:2015}. In light of the previous {\it H} band detection \citep{Currie:2015}, we investigate the source characteristics in detail and propose an alternative interpretation.

\subsection{Astrometry}
\label{sec:astro}
\begin{figure}[th]
      \centering
    \includegraphics[width=0.5\textwidth]{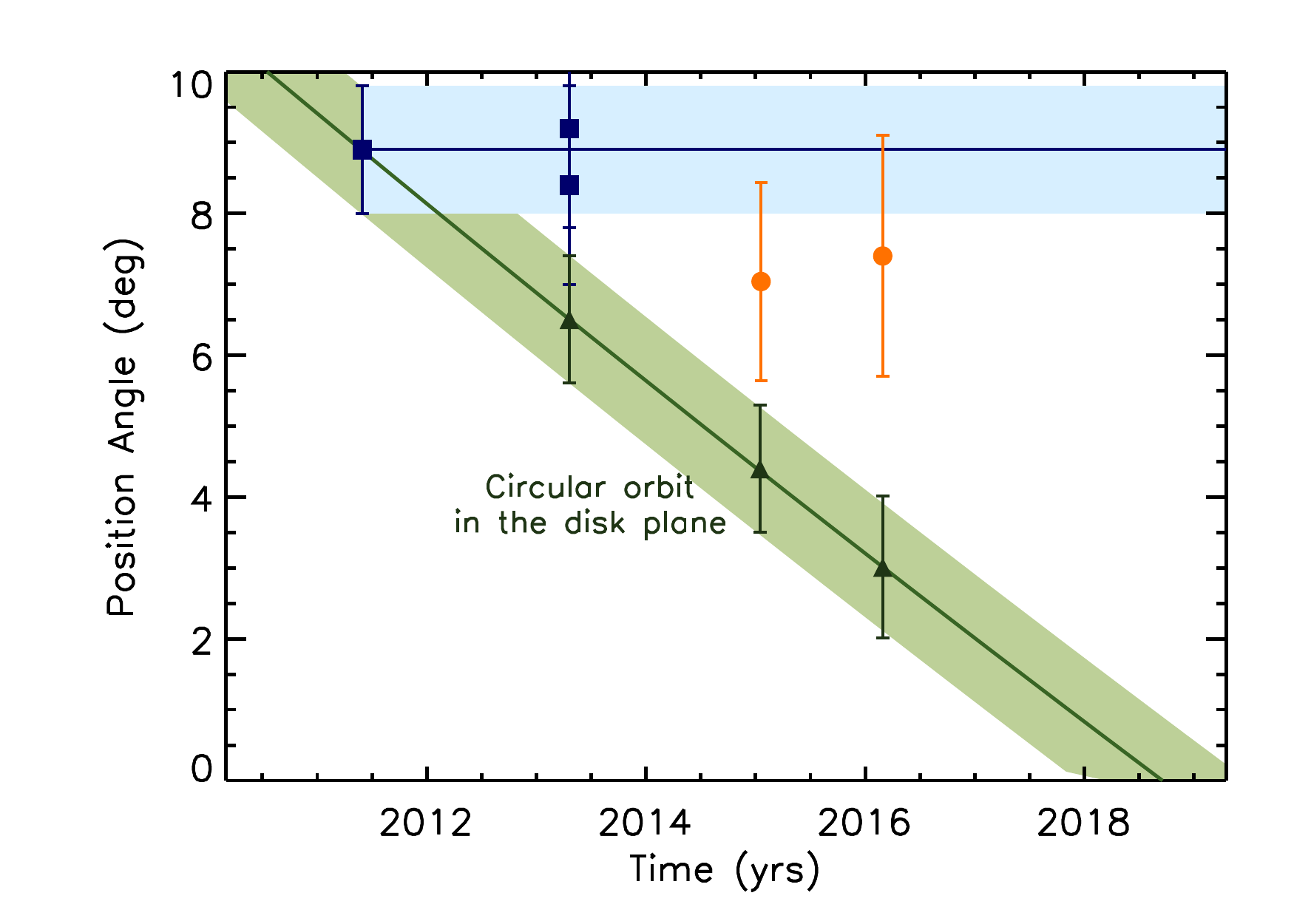}
     \caption{Measurements of the position angle of HD 100546 b with VLT/NaCo (blue squares, \citealt{Quanz:2013,Quanz:2015}) and of the detected source with GPI (orange circles, \citealt{Currie:2015} and this work). A blue line represents unchanging position between the first and the last epoch, while the projected motion, from the first epoch, following a Keplerian circular clockwise orbit in the disk plane ($i=42$ deg, $\Omega$=145 deg) is plotted in green, with green traingles representing the expected positions at each epoch. If the orbital direction is counterclockwise, then it is even less compatible with the PA reported in 2016.}
    \label{fig:astrometry}
\end{figure}

Considering that the planet is still in formation, it is fair to assume that it orbits in the disk plane, with a circular orbit. Eccentricity growth, if any, is expected to occur after the gas dissipation, which has not yet occurred in this system. With the caveat that inward migration is possible \citep{crida07}, the Keplerian motion of HD 100546 b at a deprojected distance of $59\pm2$ au around the $2.4\pm0.1$ \Ms central star is $1.4\pm0.1$ deg/yr, thus $6.8\pm0.4$ deg projected on the disk plane between the first detection in June 2011 and our observations in February 2016. Using the position angle of $8.9\pm0.9$ deg from \citet{Quanz:2013}, HD 100546 b should be $3.0\pm1.0$ deg North East of the star in 2016 if rotating clockwise or $14.8\pm1.0$ deg if rotating counterclockwise since its rotation direction is unknown.

The best fit astrometry of the source in our GPI data gives a projected separation of $471\pm9$ mas and a position angle of $7.4\pm1.7$ deg, where the uncertainties are combined in quadrature from the errors of the measurement, the star center, the plate scale, and the position angle offset. These measurements are consistent with those of \citet{Currie:2015}.

All of the reported projected separations remain unchanged to within $1\sigma$ over the four year time-span, which is consistent with both the stationary or a Keplerian orbit (only $\approx14$ mas of total decrease). However, if the {\it H} band emission corresponds to a physical objet, then its position angle is $1.9\sigma$ away from the expected position assuming clockwise rotation of the system, and $3.7\sigma$ away from the counterclockwise prediction. 

Therefore, the location of the source seen in our GPI data is consistent with an absence of motion within 1 $\sigma$. (see Figure \ref{fig:astrometry}).  

Non-Keplerian orbital motion might be expected for HD 100546 b if the planet undergoes migration through disk-planet interaction \citep{crida07,Baruteu16}. Given the estimated mass of the planet (several Jupiter masses), classical inward type II migration would be expected on timescales of $10^4-10^5$ yrs. However, this scenario predicts a fast decrease of the separation of the protoplanet and a cleared gap (a necessary condition for fast type II migration) at levels that would likely already have been detected (see \ref{sec:accretion} for further details on the gap non-detection). The migration rate and direction can be significantly altered if the gap is only partially cleared due to a high viscosity of the disk  \citep{Baruteu16}. Dramatic slow-down of the migration rate or even slow outward migration cannot be rejected based on the astrometry collected so far.

\subsection{Emission}
\label{sec:spectrum}
\begin{figure*}[th] 
\centering
\includegraphics[scale=0.55,angle=-90,clip,trim=3.5cm 0cm 4cm 2cm]{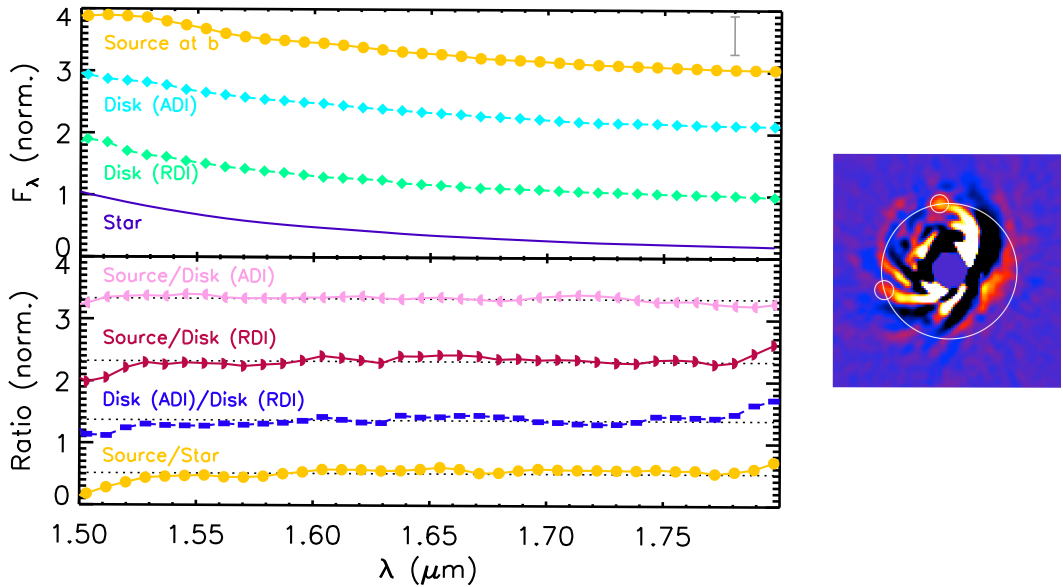} 
\caption{H band normalized spectra of the source at the expected location of HD 100546 b (orange circles) and of the disk at the same angular separation but different position angle in the ADI reduced image (light blue diamonds) and at the same disk location in the RDI reduced image (mint diamond) (see image inset). A NextGen model of the star ($T_\mathrm{eff}=10500$ K, \citealt{vanderancker97}) at the resolution of GPI is also plotted (purple line). All spectra have been normalized and arbitrarily offset for clarity. The typical measurement uncertainty is shown by the vertical grey line (top left), measured at the same angular separation in a disk-free region of the images. (Bottom) Corresponding ratios of the spectrum of the source over that of the ADI-reduced disk (magenta left half circle), of the source over that of the RDI-reduced disk (crimson right half circle), of the two disk reductions (blue bar), and of the source over that of the star, i.e., contrast (orange circle). These ratios emphasize the pure scattered light emission of the source and the absence of ADI bias.}
    \label{fig:spectr}
\end{figure*}

\begin{figure}[th]
      \centering
    \includegraphics[width=0.5\textwidth]{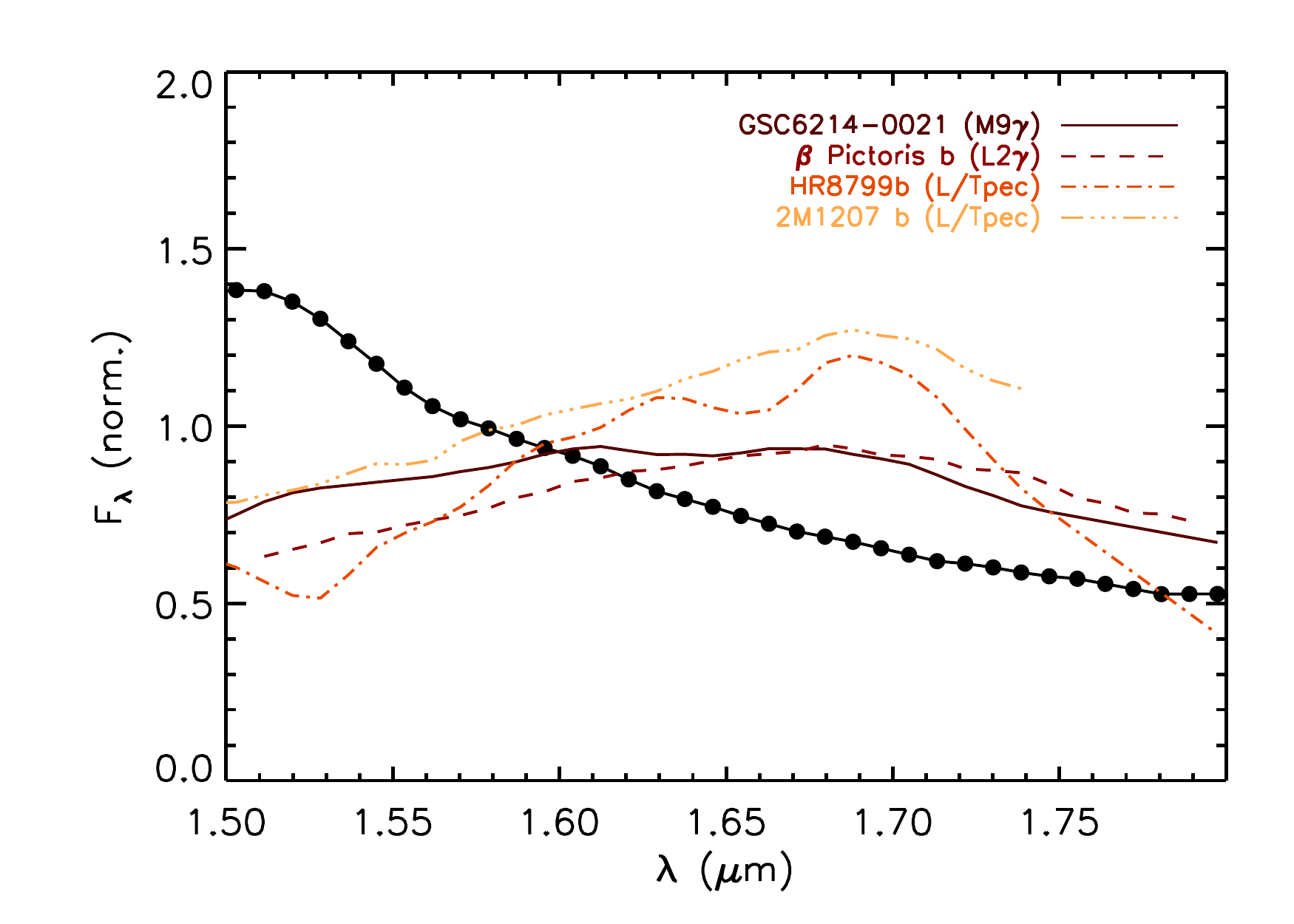}
     \caption{Spectrum of the source at the expected location of HD 100546 b (black circles) compared to spectra of young substellar objects: GSC6214-0021 (brown solid line, \citealt{lachapelle15}), $\beta$ Pictoris b (dashed red line, \citealt{Chilcote17}), HR8799 b (dashed dotted orange line, \citealt{Barman:2011}), and 2M1207 b (light orange dashed triple dotted line, \citealt{Patience:2010}). These spectra are binned to the resolution of GPI in {\it H} band (45) and scaled to the spectrum of the source by their integrated flux. Flux units are arbitrary. Typical error of the spectrum is of the order of $14\%$ of the flux.}
    \label{fig:comp_spec}
\end{figure}

If the source detected in our GPI data is indeed the protoplanet HD 100546\,b with an estimated effective temperature of $\simeq900 $K \citep{Quanz:2015}, its {\it H} band spectrum should be different from that of the surrounding disk. It should exhibit either water absorption if it resembles planets like HR 8799 b \citep[e.g.,][]{Barman:2011}, or a very red slope without molecular absorption if it is very dusty like the hotter 2M1207 b \citep{Patience:2010}. Alternatively, if the protoplanet is accreting and the emission mostly originates from the circumplanetary disk, the {\it H} band spectrum should have either a red slope or a triangular shape, depending on the accretion rate and the inner disk radius \citep{Zhu:2015}. The best fit contrast of the source in our GPI data is $\Delta H=13.2\pm 0.3$ mag, where the uncertainties are combined in quadrature with the errors from the measurement and the star-to-spot ratio \citep{Maire:2014gs}. Assuming the source is embedded in the disk, the dust extinction at this location is $A_H=3.4$ mag \citep{Currie:2015}, leading to $H=15.8\pm 0.3$ mag, and $H-L\,'=3.1\pm 0.3$ mag. These values are consistent with the predictions of an accreting protoplanet with $M_\mathrm{p}\dot{M}=3\times10^{-6}$ $M^2_\mathrm{Jup}.yr^{-1}$ and an inner disk radius of $1$ $R_\mathrm{Jup}$ \citep{Currie:2015}. Therefore, the predictions for the source from \citet{Zhu:2015} suggest a red featureless {\it H} band spectrum.

Our GPI data can test the above scenario by comparing the spectrum of the source with that of the disk. The most aggressive $4$ pixels high-pass filtered PCA reduction was used for that purpose (Fig. \ref{fig:gpiimage}, left) since it shows the resolved source and many bright disk structures. Like the astrometry, the contrast of the source was extracted in each wavelength channel at the best fit position and FWHM ($6.8$ pixels). For comparison, contrasts of the disk were also extracted, using aperture photometry with a diameter of $6.8$ pixels, from a location South East of the star at the same projected separation. The regions of these extractions are shown in Figure \ref{fig:spectr}. Since ADI might bias the extracted disk contrasts, they were extracted at the candidate and opposition disk locations in the RDI reduced residual wavelength channels. All contrasts were then multiplied by the spectrum of the $10500$ K central star \citep{vanderancker97}, which was obtained by averaging BT-NextGen models \citep{Allard:2009} at $10400$ and $10600$ K and binning to the resolution of GPI.

To estimate the uncertainties of the measurements, each ADI wavelength channel image was convolved by the same aperture. The standard deviation of the pixels at the separation of the extractions was then taken in a region free of astrophysical signal, giving the final error. Typical noise is $14\pm3\%$ of the source contrast across the band. The same exercise in the RDI reduced images provides typical $18\pm2\%$ typical noise. 

We do not flux calibrate the spectra but simply perform a comparison by eye since only the relative slopes and absorption features are of interest. Indeed, calibrating the self-subtraction of the disk due to ADI requires precise modelling \citep[e.g.,][]{Esposito:2014} that is beyond the scope of this paper. Nevertheless, since the reduction parameters are the same for all wavelength channels, ADI self-subtraction should be achromatic and only a function of the separation, which is why the measurements were done at the same angular distance.

The final spectra are plotted in Figure \ref{fig:spectr}. The {\it H} spectrum of the source at the location of HD 100546 b exhibits no absorption feature and has a blue slope. The comparison with the star spectrum proves the detected emission is pure scattered light as seen when compared to  the disk spectrum at the same separation. The spectrum is also compared to that of several young substellar objects with similar mass in Figure \ref{fig:comp_spec}: GSC6214-0021 (M9$\gamma$, \citealt{lachapelle15}), $\beta$ Pictoris b (L2$\gamma$, \citealt{Chilcote17}), HR8799 b (L/Tpec, \citealt{Barman:2011}), and 2M1207 b (L/Tpec, \citealt{Patience:2010}). These spectra are binned to the resolution of GPI in {\it H} band and interpolated over the same wavelength grid. This comparison further emphasizes the dissimilarities with the emission from a young planet.

This pure stellar-like spectrum is therefore inconsistent with a planetary photosphere, dust-obscured companion, or accreting circumplanetary disk source.

\subsection{Accretion}
\label{sec:accretion}
\begin{figure*}[th]
    \centering
    \includegraphics[width=0.8\textwidth,clip,trim=0 2cm 0 3cm]{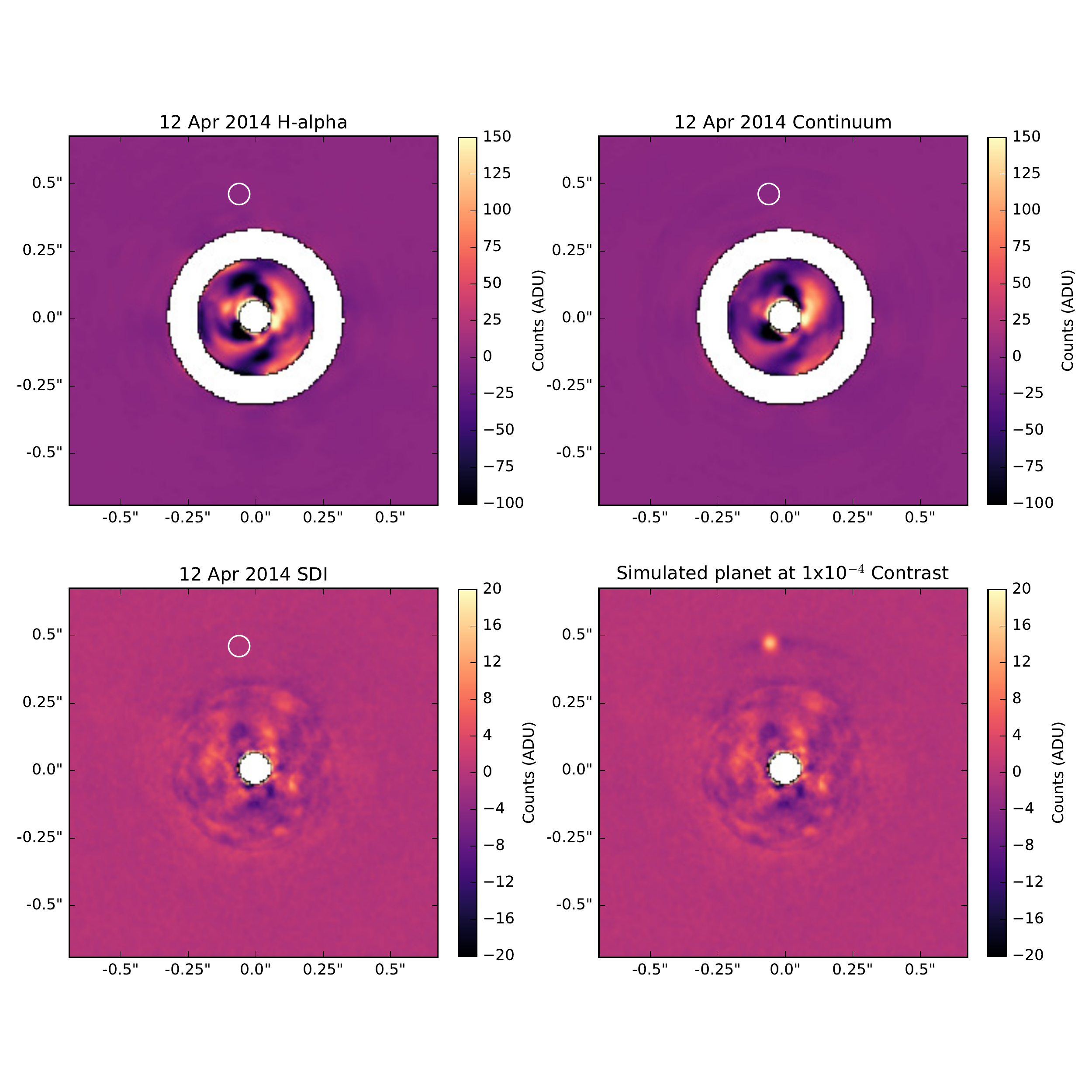}
    \caption{KLIP-processed MagAO {\it H}$\alpha$ (upper left) and continuum (upper right) images. Individual {\it H}$\alpha$ and continuum images are scaled to generate SDI images with KLIP, which are also processed with KLIP (lower left). No point sources are visible in this reduction, including at the location of ``b" (white circle in all images). The lower right image contains a simulated planet injected at a contrast of $\sim$1$\times$10$^{-4}$, equivalent to the computed 5$\sigma$ limit at this location. The white zone is used to mask out the region near the control radius of the adaptive optics where significant residual noise remained after the PSF subtraction processing.}
    \label{fig:magao}
\end{figure*}

MagAO {\it H}$\alpha$ images in Figure \ref{fig:magao} reveal structure in the inner $\sim$0$\farcs$2 of the disk, and nothing of note outside of this radius. The signal in the inner region is clearly scattered light, as scaling by the stellar {\it H}$\alpha$/continuum ratio and subtracting creates structure-free SDI images. When these SDI images are KLIP processed, no {\it H}$\alpha$ excess is apparent at the location of HD\,100546\,b.

Correcting the 5$\sigma$ ASDI contrast limit at this radius for KLIP throughput reveals that the maximum {\it H}$\alpha$ contrast at the location of HD\,100546\,b is $\sim$1$\times$10$^{-4}$, under the assumption that no flux is present in the continuum channel. The contrast limit at this location can be used to place upper limits on the accretion properties of the protoplanet, given several assumptions as described below. The non-detection of an {\it H}$\alpha$ source at this location is indeed not necessarily inconsistent with the existence of an accreting protoplanet.

Accreting protoplanets, and giant planets in particular, are expected to clear circular regions in the disk, each several Hill radii in extent (about 7 au in our case) \citep{dodson:2012} and extending azimuthally as they orbit. Therefore, the protoplanet should lie in a region free of extinction by the dust disk and the accretion luminosity would be directly detectable down to the sensitivity limit. Following \citet{close14} and \citet{Sallum:2015ej}, we use the contrast limit in {\it H}$\alpha$ to derive an upper limit on the accretion luminosity onto the protoplanet. Assuming the extinction towards the protoplanet is the same as the star, we derive $A_R=0.11$ mag from $A_V=0.15$ mag \citep{satori03} and the standard interstellar dust extinction law \citep{Cox:2000}. Then, given the width and zero point of the {\it H}$\alpha$ filter, the distance and magnitude of HD\,100546, $A_R$, and the contrast limit, we derive $\mathrm{L}_{\mathrm{H}\alpha}<1.4\times10^{-5} ~\mathrm{L}_\odot$ and, considering that accretion laws from T-Tauri stars also apply to planetary mass objects \citep{Rigliaco12}, ~$\mathrm{L}_\mathrm{acc}<1.7\times10^{-4} ~\mathrm{L}_\odot$. The accretion rate $\dot \mathrm{M}$ can be computed from the accretion luminosity with the mass ($\mathrm{M}$) and radius of the accreting object \citep{Gullbring:1998}. We consider radii of 1 $\mathrm{R}_\mathrm{Jup}$ and 2 $\mathrm{R}_\mathrm{Jup}$ to bracket plausible values from evolutionary models \citep{Baraffe:2003bj} but treat the mass as a free parameter as it is very uncertain. Therefore, we place upper limits on $\mathrm{M}\dot\mathrm{M}< 6.3\times10^{-7}~\mathrm{M}_\mathrm{Jup}^2.\mathrm{yr}^{-1}$ and $<1.3\times10^{-6}~\mathrm{M}_\mathrm{Jup}^2.\mathrm{yr}^{-1}$, for 1 and 2 $\mathrm{R}_\mathrm{Jup}$ respectively, under the assumption that the protoplanet is in a clear region. For comparison,  with a possible mass from a few to 15 $\mathrm{M}_\mathrm{Jup}$ for HD\,100546\,b, the accretion flow on the protoplanet, if any, remains very small in comparison to that of on the central star ($\mathrm{M}\dot \mathrm{M}\sim2\times10^{-1}~\mathrm{M}_\mathrm{Jup}^2.\mathrm{yr}^{-1}$, \citealt{Mendigutia:2015}). The upper limit on the accretion rate rules out the previously proposed scenario in which the {\it H} band emission is coming from an accreting protoplanetary disk with an inner radius of 1 $\mathrm{R}_\mathrm{Jup}$ and $\mathrm{M}\dot\mathrm{M}=3.6\times10^{-6}~\mathrm{M}_\mathrm{Jup}^2.\mathrm{yr}^{-1}$ \citep{Currie:2015}.\\
The upper limits on the accretion rate and luminosity are typical for T-Tauri stars \citep{Gullbring:1998,Rigliaco12}, or the young planet LkCa\,15\,b % with $\mathrm{L}_\mathrm{acc}=4\times10^{-4} \mathrm{L}_\odot$ and $\mathrm{M}\dot\mathrm{M}=3\times10^{-6} \mathrm{M}_\mathrm{Jup}^2.\mathrm{yr}^{-1}$ 
\citep{Sallum:2015ej}. On the other hand, the protoplanet may be temporarily in a quiet phase and remains undetectable since accretion is known to be stochastic on TTauri stars \citep{bouvier07}.

Alternatively, it has been hypothesized that HD\,100546\,b is still embedded in the disk \citep{Quanz:2013,Currie:2015}. The non-detection of any gap larger than few au around the protoplanet, despite a resolution of 2 au in the optical \citep{Garufi:2016}, does not reject the hypothesis of a deeply embedded planet. With $A_H=3.4$ mag \citep{Currie:2015}, the extinction due to the disk in {\it H}$\alpha$ is $A_R=22$ mag. Therefore, the accreting protoplanet would likely not be detectable in {\it H}$\alpha$. We can still do the same exercise and place some constraints on the accretion luminosity and rate. We find $\mathrm{L}_\mathrm{acc}<8\times10^{3}~\mathrm{L}_\odot$ and $\mathrm{M}\dot\mathrm{M}< 1.5\times10~\mathrm{M}_\mathrm{Jup}^2.\mathrm{yr}^{-1}$ for a 1 $\mathrm{R}_\mathrm{Jup}$ planet.  As expected, these upper limits place no meaningful constraint on the accretion properties of a possible deeply embedded protoplanet.

\section{Conclusions} \label{sec:conc}
HD\,100546 hosts a well-studied transition disk with a protoplanet detected in L\,' and M' band with VLT/NaCo \citep{Quanz:2013,Quanz:2015} and a recently proposed detection in {\it H} band with Gemini/GPI \citep{Currie:2015}. With new GPI and Magellan/VisAO data, we present in this paper a source detected only in {\it H} band at the expected location of HD\,100546\,b. Detailed analyses reveal that:
\begin{itemize}
    \item The source is located at the tip of a bright disk arc and non-aggressive high-pass filtering and reduction processes reveal that the source is contiguous with the underlying disk feature, similarly to blobs aligned along other spiral arms;
    \item The size of the source, point-like only in aggressive reductions, steadily decreases with that of the high-pass filter. This demonstrates that the source itself is extended;
    \item The astrometry of the source is consistent with stationary motion with respect to the discovery epoch and inconsistent at a $2\sigma$ level with a Keplerian circular orbit in the disk plane;
    \item The spectrum of the source is consistent with that of the disk and with pure scattered-light, proving the emission is not coming from a planet photosphere or an accreting circumplanetary disk;
    \item The non detection of a source at the $5\sigma$ level at this location in {\it H}$\alpha$ places upper limits, with the hypothesis the protoplanet is a cleared disk region, on the accretion luminosity of $1.7\times10^{-4} \mathrm{L}_\odot$ and on the accretion rate of and $6.3\times10^{-7}~\mathrm{M}_\mathrm{Jup}^2.\mathrm{yr}^{-1}$ for a radius of 1 $\mathrm{R}_\mathrm{Jup}$. This rules out the presence of the accreting circumplanetary disk previsouly proposed \citep{Currie:2015} but instead does not exclude the embedded hypothesis.
\end{itemize}

Therefore, the source detection in {\it H} band benefits from visual inspections of its morphology and spectrum, a 2$\sigma$ rejection of clockwise keplerian orbit, and a 5$\sigma$ non-detection of accretion light in {\it H}$\alpha$. Altogether these lines of evidence enable us to suggest that the source detected in {\it H} band with GPI is not related to HD 100546 b but is a disk feature, enhanced by the data reduction process, and the bulk of the detected flux comes from the disk. However, none of them refute the embedded hypothesis: planet-disk interaction (migration) might make the orbital motion non-keplerian and emissions in {\it H} band and {\it H}$\alpha$ from the protoplanet might be blocked by the surrounding circumstellar disk.

Considering that the detected source is pure disk emission in H-band that happens to be at the location of HD\,100546\,b, one wonders whether disk emission can contribute to the detection at longer wavelenghts. The source has $H=19.2\pm 0.3$ mag without dust extinction. From the H-L\,'$=-1.08\pm0.35$ mag color of the disk \citep{avenhaus14}, the source that we detect would have $\Delta$L\,'$=15.4\pm0.5$ mag assuming a grey albedo, compared to $\Delta$L\,'$=9.4\pm0.1$ mag of HD\,100546\,b reported by \citet{Quanz:2013}. Therefore the disk alone cannot produce the the detected L$\,'$ emission and a local source of heat is necessary to explain it. The non-detection of HD\,100546\,b in {\it H} band in our data is also consistent with the non-detection in {\it K} band in VLT/SPHERE-IRDIS data \citep{Garufi:2016}. The most logical explanation for these multiwavelength data is that the optically thick circumstellar disk could be blocking thermal emission from the deeply embedded protoplanet, preventing any detection in the visible/near-infrared. The most problematic issue is the absence of polarization enhancement coming from the disk itself at the location of the protoplanet \citep{Quanz:2013, avenhaus14,Garufi:2016}, which is hard to reconcile with the scattered-light emission seen in{\it H} band. To resolve this problem, the disk architecture and properties of the dust have to be better understood. ALMA observations at high resolution might reveal the presence of the circumplanetary disk or a peculiar morphology at the location of HD\,100546\,b. The size of such structures could be used to infer the mass of the protoplanet and help to understand disk clearing mechanisms and planet formation. JWST promises discerning thermal spectroscopy of the protoplanet that could reveal absorption features placing stronger constraints on its physical properties.

\acknowledgments

Based on observations obtained at the Gemini Observatory, which is operated by the Association of Universities for Research in Astronomy, Inc., under a cooperative agreement with the National Science Foundation (NSF) on behalf of the Gemini partnership: the NSF (United States), the National Research Council (Canada), CONICYT (Chile), the Australian Research Council (Australia), Minist\'{e}rio da Ci\^{e}ncia, Tecnologia e Inova\c{c}\~{a}o (Brazil) and Ministerio de Ciencia, Tecnolog\'{i}a e Innovaci\'{o}n Productiva (Argentina). This work has made use of data from the European Space Agency (ESA)
mission {\it Gaia} (\url{http://www.cosmos.esa.int/gaia}), processed by the {\it Gaia} Data Processing and Analysis Consortium (DPAC,
\url{http://www.cosmos.esa.int/web/gaia/dpac/consortium}). Funding
for the DPAC has been provided by national institutions, in particular
the institutions participating in the {\it Gaia} Multilateral Agreement. J.R., R.D. and D.L. acknowledge support from the Fonds de Recherche du Qu\'{e}bec. KBF and JRM's work was performed in part under contract with the California Institute of Technology (Caltech)/Jet Propulsion Laboratory (JPL) funded by NASA through the Sagan Fellowship Program executed by the NASA Exoplanet Science Institute. Supported by NSF grants AST-1411868 (K.B.F. and B.M.), AST-141378 (G.D.), and  AST-1411868 (JP). Supported by NASA grants NNX14AJ80G (E.L.N., S.C.B., B.M., F.M., and M.P.), NNX15AD95G (BM, JEW, TME, RDR, GD, JRG, PGK) and NNX16AD44G (K.M.M, T.S.B., and L.M.C.). KWD is supported by an NRAO Student Observing Support Award SOSPA3-007. Portions of this work were performed under the auspices of the U.S. Department of Energy by Lawrence Livermore National Laboratory under Contract DE-AC52-07NA27344.

{\it Facility:} \facility{Gemini:South (GPI)} and \facility{Magellan:Clay (MagAO)}

\end{document}